\documentstyle{article}

\author{Christof Zalka}

\begin{document}
\hfill BUTP--96/11
{\center \bf \LARGE Efficient Simulation of Quantum Systems \\
         by Quantum Computers \\}
\vspace{0.5cm}
 {\center \large Christof Zalka \\}
{\center \it Institut f\"ur theoretische Physik, Universit\"at Bern,
Switzerland \\} 
{\large \center March 25, 1996 \\}
\vspace{0.5cm}

\begin{abstract}
We show that the time evolution of the wave function of a quantum mechanical
many particle system can be implemented very efficiently on a quantum
computer. The computational cost of such a simulation is comparable to the
cost of a conventional simulation of the corresponding classical system. We
then sketch how results of interest, like the energy spectrum of a system, can
be obtained. We also indicate that ultimately the simulation of quantum field
theory might be possible on large quantum computers.

We want to demonstrate that in principle various interesting things can
be done. Actual applications will have to be worked out in detail also
depending on what kind of quantum computer may be available one day... 
\end{abstract}

\section{Quantum Computers (QCs)}

Quantum computers are still imaginary devices, but it is hoped that eventually
the technical problems involved in their realization can be overcome
\cite{Lloyd,Chuang,Yamamoto,Laflamme,Barenco,Barenco2,Unruh}. Quantum
computers could solve some problems much faster than conventional
computers. Most prominently, Peter Shor (1994) has given a "quantum algorithm"
for factoring large integers in polynomial time \cite{Shor}(see also
\cite{Chuang}).

A $l$-bit quantum computer may be thought of consisting of $l$ two-state
systems. Computations would be carried out by inducing unitary transformations
of a few at a time of these quantum-bits (qubits). This may be done by
exterior fields controlled from the outside. Upon observation at the end of
the "unitary computation" the quantum computer would collapse into a state
where each qubit is either a 0 or a 1. The art of finding a "quantum
algorithm" is to extract useful information from a few such runs.

The main technical problem in realizing QCs is, to prevent unwanted
interactions with the environment during the calculation. Such interactions
can cause decoherence, such that the QC is no longer in a pure quantum
state. Here we consider an idealized QC which doesn't suffer from such
problems.

\section{Simulating Quantum Systems}

General ideas about using specially designed quantum systems to simulate other
quantum systems have been published, e.g. by Feynman \cite{Feynman}. I
present here an actual implementation of the simulation of quantum mechanical
many particle systems on a general purpose QC \footnote{Independently and at
practically the same time as me, Steven Wiesner \cite{Wiesner} has published a
preprint concerning this, using a slightly different approach}. For every
degree of freedom of the system we need a $l$-bit quantum register. This
allows to store the whole (discretized) wave function as amplitudes of the
"classical" states in the quantum computer. Already to store so much
information would be impossible for a conventional computer. It is then shown
that the time evolution for short time steps amounts essentially to a Fourier
transformation which can be carried out very efficiently on a quantum computer
\cite{Shor,Coppersmith,Ekert}.

\subsection{Quantum Mechanical Particle in 1 Dimension}

First we discretize the wave function and impose periodic boundary conditions:

\begin{equation} a_n = \psi(n~\Delta x) \qquad a_{n+N}=a_n \quad . 
\end{equation}
After proper normalization we store these amplitudes in a $l$-bit quantum
register:

\begin{equation} |\psi \rangle = \sum^{N-1}_{n=0} a_n |n \rangle \qquad
 N=2^l \quad . \end{equation} 
Where $|n\rangle$ is the basis state corresponding to the binary
representation of the number $n$. For short time steps $\Delta t$ the Greens
function of the Schr\"odinger equation is approximately:

\begin{equation} G(x_1,x_2,\Delta t) = k~ e^{i\frac{m}{2}      \label{Green}
\frac{(x_1-x_2)^2}{\Delta t} - i V(x_1) \Delta t}  \quad . \end{equation}
Applying this to the amplitudes $a_n$ is equivalent to transforming the basis
states with the inverse transformation. Using proper normalization to make the
transformation unitary, we get:

\begin{equation} |n\rangle \to \frac{1}{\sqrt{N}} \sum^{N-1}_{n'=0}
 e^{-i\frac{m}{2} \frac{(n-n')^2 {\Delta x}^2}{\Delta t} 
 + i V(n~ \Delta x) \Delta t}~  |n'\rangle  \quad . \end{equation} 
The crucial observation is that this corresponds to two diagonal unitary
matrices plus a discrete Fourier transformation:

\begin{equation} |n\rangle \to \underbrace{\frac{1}{\sqrt{N}} 
e^{-i\frac{m}{2} \frac{n^2 {\Delta x}^2}{\Delta t}+i V(n~ \Delta x) \Delta
t}}_{\mbox{diagonal}}~ 
\underbrace{\sum^{N-1}_{n'=0} e^{i m \frac{n n' {\Delta x}^2}{\Delta
t}}}_{\mbox{Fourier trafo}} 
\underbrace{\left( e^{-i\frac{m}{2} \frac{n'^2 {\Delta x}^2}{\Delta
t}} |n'\rangle \right) }_{\mbox{diagonal}} \quad . 
\end{equation}
To be able to use the fast Fourier transformation algorithm (FFT), we chose the
parameters such that: 

\begin{equation} m \frac{{\Delta x}^2}{\Delta t} = A \frac{2 \pi}{N}~,
 \qquad \qquad A \quad \mbox{integer}  \quad . \end{equation}
Provided the potential $V$ has been made periodic (with period N), this also
makes the whole phase factor periodic in $n$ and $n'$, as is necessary for
consistency. It is rather straightforward to implement the FFT algorithm on a
quantum computer \cite{Shor,Coppersmith,Ekert} such that it uses only about
$l^2/2$ computational steps (local unitary transformations). The
diagonal unitary transformations of the type

\begin{equation} \label{rephasing} |n\rangle \to e^{i c F(n)}~ |n\rangle  
\quad , \end{equation}
can be done with the following succession of steps:

\begin{equation} \label{phase} |n,0\rangle \to |n,F(n)\rangle \quad ; \qquad
|F(n)\rangle \to e^{i c F(n)}~ |F(n)\rangle \quad ; \qquad |n,F(n)\rangle \to
|n,0\rangle \quad . \end{equation}
In the first and last step the vectors represent two quantum registers. The
first step corresponds to parallel classical calculations. It can be carried
out on a quantum computer with a somewhat higher cost than just one such
calculation on a conventional computer (P. Shor
\cite{Shor}). As all transformations are unitary they can just as well be
inverted, so that the last step is also possible. The second step is really a
transformation of the form

\begin{equation} |n\rangle \to e^{i c n} |n\rangle  \quad , \end{equation}
which can be carried out in a straightforward manner in $l$ steps by
transforming each bit separately by an appropriate transformation. Note that
of course all these operations are carried out in parallel on all basis states
which is why this is often called "quantum parallelism".

\subsection{Many Particles and Field Theory}

The generalization to many particles is straightforward. For $n$ particles in
3 dimensions we need $3 n$ quantum registers. Besides the above steps we need
transformations corresponding to the coupling between particles and between
the different degrees of freedom of the same particle. They are diagonal
unitary transformations acting on several registers, e.g.:

\begin{equation} |n,n',n''\rangle \to e^{i c F(n,n',n'')}~ |n,n',n''\rangle  
\quad . \end{equation}
Such a transformation can be carried out analogously to eq.(\ref{phase}).
For identical particles the initial state of the quantum computer has to be
chosen symmetric resp. antisymmetric. Also the generalization to more
general Hamiltonian operators than in the previous paragraph is possible (e.g.
with coupling to a magnetic field).

Also quantum field theory could be simulated, e.g. by discretizing the field as in
lattice gauge theory (for a standard text see e.g. \cite{Stulz}). For a bosonic quantum field this would be equivalent to
having 1 particle at each lattice point. At any rate, applications like lattice
QCD would require large quantum computers with thousands of quantum bits, which
are not going to be realizable soon.

\subsection{Fermionic Field Theories}

Fermionic quantum fields pose some problems, as their functional formulation
(path integrals, wave functionals,...) involves anticommuting Grassmann
numbers. There is no wave functional in terms of the usual complex numbers, so
we have to look for a representation of the fermionic field operator-algebra
on some other Hilbert space. One possibility is the Fock space, where we have
for the $n$-particle sector a totally antisymmetric wave function of
$n$ variables. Another possibility is to give the occupation number for each
particle species at each lattice point. For fermions this can only be 0 or 1
and thus requires only 1 qubit. 

I expect, that further difficulties which may arise with quantum fields can in
principle be overcome, possibly along the same lines as it is done in lattice
QCD. 

\section{Other manipulations}

\subsection{Simulating a decay to obtain the ground state}

Often one is interested in the ground state of a quantum system, be this the
ground state of a molecule, the vacuum of a field theory or a hadron stable
under the strong interaction. I propose here to obtain the ground state
essentially in the same way as it happens in reality, namely by letting some
initial state decay. Decays happen because the instable particle is coupled to
some other quantum system, like the electromagnetic field. It is important
that this quantum system has a lot of different energy eigenvalues such that
it can absorb the energy from various transitions in the decaying system. For
the simulation of a quantum computer we may take a collection of 2-level
systems with energy gaps $\Delta E = E_0~ 2^{-n}, ~n=0 \dots l~$ such that the
whole system has practically a continuous energy spectrum. The coupling to the
quantum system of interest may be done in various ways.

During the simulation of the decay, the auxiliary "energy-drain" system must
periodically be reset to its ground state such that it is able to absorb more
energy. As the study of simple systems shows, this does not necessarily in
every step, but only in the average, lead to a reduction of the energy
expectation value of the system of interest. For actual implementations one
may have to think about an adequate "resetting" strategy, e.g. depending on
the state the auxiliary system is found in upon observation just before each
resetting. Also one may want to reduce the coupling towards the end of the
decay simulation in order not to disturb the ground state of interest too
much.

Actually it may be enough to simulate the decay until the few lowest lying
states (which are of interest) have a large amplitude. Then one can follow the
procedure of paragraph \ref{measure} to obtain one of the energy eigenstates.

\subsection{Putting a wave function on a Quantum Computer}

Here I sketch a method which in principle allows us to store a given wave
function as an initial state on a QC. I demonstrate this for the case of a
scalar quantum mechanical particle in 1 dimension. The generalization to multi
dimensional wave functions is then rather straightforward. Here I show only
how the absolute value of the wave function can be stored in a quantum
register, as above I have already shown how the complex phase can later be
introduced.

So we want to carry out the following transformation on a $l$ bit quantum
register which initially is in the state $|0\rangle$:

\begin{equation}  |0 \rangle \to k ~\sum^{2^l-1}_{n=0} \left| \psi \left( n 
\frac{L}{2^l} \right) \right| ~|n\rangle \quad , \qquad 
\mbox{where} \quad \psi(x+L)=\psi(x)\quad , 
\end{equation}
and $k$ is an appropriate normalization constant. The idea now is to split the
norm of the initial state (which is 1, of course) $l$ times such that in the
end we get the $2^l$ contributions for the single basis states
$|n\rangle$.\footnote{Note that the norm of mutually orthogonal substates is a
useful notion in quantum computing because it is conserved under unitary
transformations.} (Imagine a binary tree with $l$ levels and $2^l$ leaves.)
For this to work out correctly each split has to yield the right ratio. For
this we must have an algorithm which can compute the integrals

\begin{equation}  I_{i,k}= \int^{\frac{k+1}{2^i}L}_{\frac{k}{2^i}L}
|\varphi(x)|^2 dx \quad , \qquad \mbox{for} \quad k=0 \dots 2^i-1 \quad
\mbox{and}  \quad i=1 \dots l \quad . \end{equation}
Now the first split is realized by a real O(2) rotation acting on the most
significant qubit of the quantum register, thus:

\begin{equation}  |0,\dots \rangle \to \sqrt{I_{1,0}}~ |0,\dots \rangle
         +  \sqrt{I_{1,1}}~ |1,\dots \rangle \quad . \end{equation}
Such an O(2) rotation of a qubit can be carried out analogously to the
rephasing of an amplitude (see eq. (\ref{rephasing})) once the rotation angle
$\varphi$ has been calculated:

\begin{equation} \sin \varphi = \sqrt{I_{1,1}} \quad . \end{equation}
The subsequent calculations of angles and rotations are then carried out in
quantum parallelism, thus:

\begin{equation} |b_l,b_{l-1},...b_{l-i},0,...\rangle \to
 \cos(\varphi)~ |b_l,b_{l-1},...b_{l-i},0,...\rangle +
 \sin(\varphi)~ |b_l,b_{l-1},...b_{l-i},1,...\rangle  \quad , \\
\end{equation}
\qquad \qquad \qquad with $~\sin(\varphi)=\sqrt{I_{l-i+1,k}}~$ and 
$~k=b_{l-i}+2~b_{l-i+1}+...+2^i~b^l~$. \\ \\
So this happens simultaneously for all values of the $b$'s (resp. of $k$).

The method given here to input a wave function to a QC is rather ambitious, as
essentially it evaluates the function at \underline{every} (discretized) point
$x$. In practice a simpler sceme may do.

\section{Obtaining Results of Interest}

\subsection{Measuring positions of particles, resp. field strengths}
One can think of various ways how to obtain quantities of interest from a
quantum computer simulation of a quantum system. Say we have some state on the
QC which we want to analyze, like the vacuum or a 1-particle state of a field
theory. One can now make a measurement of some field strength simply by
observing the corresponding quantum register. By repeatedly observing some
field values in different points one may in several runs of the QC get a
statistical estimate of $n$-point functions. In particular, one may obtain the
correlation length. To get $n$-point functions for space-time points with a
timelike separation, one has to continue the simulation of the time evolution
between observations of single quantum registers.

\subsection{Measuring arbitrary observables}       \label{measure}
To study the measurement problem in quantum theory, J. von Neumann
has proposed an idealized simple interaction \cite{Neumann,DeWitt} between the
quantum system and the part of the measuring apparatus that directly interacts
with the system. For convenience this part of the apparatus can be chosen to
be equivalent to a quantum mechanical particle in 1 dimension or, more
precisely, its Hilbert space is spanned by the basis vectors

\begin{equation} |x \rangle \qquad x \quad \mbox{real, and} \qquad 
\hat X |x\rangle = x |x\rangle \quad . \end{equation}
The time evolution during the measurement of an observable $\hat A$ shall then
be given by

\begin{equation} \hat U(t) |\Psi_a\rangle |x\rangle =
|\Psi_a\rangle |x+k\cdot a\cdot t\rangle \quad , \qquad \mbox{with} \quad
\hat A |\Psi_a\rangle =a |\Psi_a\rangle \quad , \end{equation}
where $|\Psi_a\rangle$ is the state of the system. One easily verifies, that
the Hamilton operator $\hat H = k \hat P \hat A$ leads to this time evolution.
Here $\hat P$ is the usual momentum operator for the imaginary quantum
mechanical particle, thus $[\hat X,\hat P]=i \hbar~$.

Now we simply have to implement this time evolution on the QC. This can be
expected to work out for any reasonable observable represented by the
hermitian operator $\hat A$ which can be written in terms of the fundamental
observables and their conjugate momenta. 

Then, after simulating this ``1. stage von Neumann measuring process'', we can
observe the auxiliary quantum register and should find a number proportional
to an eigenvalue of $\hat A$. By repeating this, we get the spectrum of $\hat
A$ and the weight of the various $\hat A$-eigenstates in the original system
state. Also after each observation the system will be in an eigenstate of
$\hat A$. Thus this is a way of obtaining e.g. the energy eigenstates.

\section{Conclusion}
I have demonstrated that on a large enough quantum computer various quantum
theoretic quantities of interest could be calculated, which may be too hard to
compute by a conventional computer. This is in particular true for strongly
interacting field theories like QCD. However simulating quantum fields would
require large quantum computers with many thousand qubits, which we don't know
whether they can ever be built. Furthermore, here I have assumed to have an
ideal QC which doesn't suffer from decoherence and can carry out unitary
transformations with sufficient precision (as these are really analog
operations).

\end{document}